\newcommand{\PRL}[3]{Phys.\ Rev.\ Lett.\ {\bf #1},\ #2 (#3)}
\newcommand{\RMP}[3]{Rev.\ Mod.\ Phys.\ {\bf #1},\ #2 (#3)}

\newcommand{\PRA}[3]{Phys.\ Rev.\ A\ {\bf #1},\ #2 (#3)}

\newcommand{\appsection}[1]{\let\oldthesection\thesection
  \renewcommand{\thesection}{ \oldthesection}
  \section{#1}\let\thesection\oldthesection}
\documentclass[twocolumn,preprintnumbers,amsmath,amsfonts,amssymb,notitlepage,superscriptaddress,showpacs,altaffillsymbol,pra]{revtex4-1}
\usepackage{geometry}
\usepackage{color}
\date{\today}
\DeclareMathAlphabet{\mathpzc}{OT1}{pzc}{m}{it}
\usepackage[utf8]{inputenc}
\usepackage{amsmath}
\usepackage{amsfonts}
\usepackage{amssymb}
\usepackage{graphicx}
\usepackage{nicefrac}
\usepackage{amsbsy}
\usepackage{float}
\usepackage{epstopdf}
\usepackage{dsfont}
\usepackage{siunitx}
\usepackage{lipsum}
\usepackage{graphicx}

\def \be{\begin{equation}}
\def \ee{\end{equation}}
\def \ba{\begin{array}}
\def \ea{\end{array}}
\def \bea{\begin{eqnarray}}
\def \eea{\end{eqnarray}}

\begin{document}
\title{Dynamics of Bose-Einstein Recondensation in Higher Bands}
\author{Vaibhav Sharma}
\thanks{vs492@cornell.edu}
\affiliation{Laboratory of Atomic and Solid State Physics, Cornell University, Ithaca, NY 14853}
\author{Sayan Choudhury}
\thanks{sc2385@cornell.edu}
\affiliation{Department of Physics and Astronomy, University of Pittsburgh, Pittsburgh, PA 15260}
\author{Erich J Mueller}
\thanks{em256@cornell.edu}
\affiliation{Laboratory of Atomic and Solid State Physics, Cornell University, Ithaca, NY 14853}

\begin{abstract}
Motivated by recent experiments, we explore the kinetics of Bose-Einstein condensation in the upper band of a double well optical lattice. These experiments engineer a non-equilibrium situation in which the highest energy state in the band is macroscopically occupied. The system subsequently relaxes and the condensate moves to the lowest energy state. We model this process, finding that the kinetics occurs in three phases:  The condensate first evaporates, forming a highly non-equilibrium gas with no phase coherence.  Energy is then redistributed among the noncondensed atoms.  Finally the atoms recondense.   We calculate the time-scales for each of these phases, and explain how this scenario can be verified through future experiments.
\end{abstract}

\maketitle

\section{Introduction}
The kinetics of ordering is one of the iconic problems in physics, with relevance to areas as diverse as cosmology and metallurgy \cite{chaikinlubensky,GuntonReview,BrayPRE94,BrayReview,Crestkinetics1,Crestkinetics2}.  New tools have evolved in cold atom systems which enable the controlled study of ordering, and which are yielding novel ordering scenarios \cite{VengalattoreReview,AltmanReview,GogolinNatp,GogolinRPP}. Recent experiments at MIT\cite{KetterleHigherBand,KetterleHigherBand2} and Hamburg\cite{hemmerichNATP,HemmerichPRL2011,HemmerichPRL2015} have observed non-equilibrium Bose-Einstein condensation in the first excited band of a bipartite optical lattice.  Similar physics is seen in Floquet lattices \cite{ChinScience2016,ChinFloquetNew,Chin2,Chin3}.  Motivated by these experiments, we study the dynamics of bosons which are condensed in the highest energy state of the first excited band of a double well optical lattice.  The system subsequently evolves to a Bose-Einstein Condensate (BEC) in the lowest energy state of that band.  We model this process, finding that the condensate first evaporates, then recondenses. This paradigm is very different from those traditionally used to model order parameter dynamics, and should have broad impact on understanding other non-equilibrium systems.  \\

Beyond their intrinsic intellectual merit, these non-equilibrium experiments are motivated by attempts to produce exotic states of matter.  The final state in the MIT experiment displays a supersolid stripe phase \cite{KetterleHigherBand,KetterleHigherBand2}.  Other higher band geometries produce even more exotic physics, ranging from multi-flavor and multi-orbital Hubbard models \cite{hemmerichNATP,HemmerichPRL2011,HemmerichPRL2015,BlochHigherBandPRL2007,WuLiuPRA,IssacsonGirvin2005,KuklovPRL2006,FisherXuPRB2007,ScarolaSDSPRL2005} to the formation of interaction-induced chiral order related to p-wave superconductivity \cite{hemmerichsmith2013,hemmerichsmith2016} or chiral Bose liquids \cite{VincentLiuNatComm2014}. A recent experiment has demonstrated the presence of a dynamical sliding phase, when P-band bosons are loaded in an one-dimensional optical lattice.\cite{slidingphase} One needs at least a qualitative understanding of the higher-band kinetics before one can reliably design protocols for producing these states.   
\

The model we use for analyzing the higher band kinetics can also be applied in other settings, including the simpler case of a Bose-Einstein condensate in the lowest band of an optical lattice.  In that case, the analogous stating point would be when all of the atoms are condensed in the highest energy state of the band.  This could be arranged by using Raman lasers, or an external force.  Alternatively, as shown in \cite{Pisa}, one can ``shake" the lattice to induce an inverted floquet band structure.  To keep our narrative as simple as possible, we will focus on the upper-band case, which motivated our study.  Limited discussion of timescales in these other experiments is given in Section~\ref{upperbandsec}. Some related theoretical work has also been done by Garcia et al~\cite{garcia} for a different model where they study the coherent dynamics and fragmentation of a BEC in a single double well potential with three modes that is quenched to a superposition state of ground and first excited mode.\\

Our paper is organized as follows. In section II, we introduce a model for analyzing the dynamics of a BEC loaded in a double well optical lattice. 
In section III, we use thermal equilibrium arguments to determine the 
properties of the system for $\tau_{N}\ll t\ll \tau_{ab}$, where $\tau_{N}$ is the microscopic scattering time in the higher band and $\tau_{\rm ab}$ is the time for decay from the upper to lower band.  In section IV and V, we describe the kinetics of condensation in the excited band, calculating $\tau_{N}$ and exploring the other timescales in the dynamics. In section VI, we calculate $\tau_{ab}$ and verify that $\tau_{\rm ab}\gg\tau_{N}$, guaranteeing that one can produce a metastable condensate in the excited band. In section VII, we discuss how time-of-flight images can be used to observe the dynamics of higher band bosons and finally in section VIII, we discuss how our model may be applied to experimental settings beyond higher bands, such as inverted bands.

\section{Model}
\subsection{Single Particle Hamiltonian}
\begin{figure}
\includegraphics[width=\columnwidth]{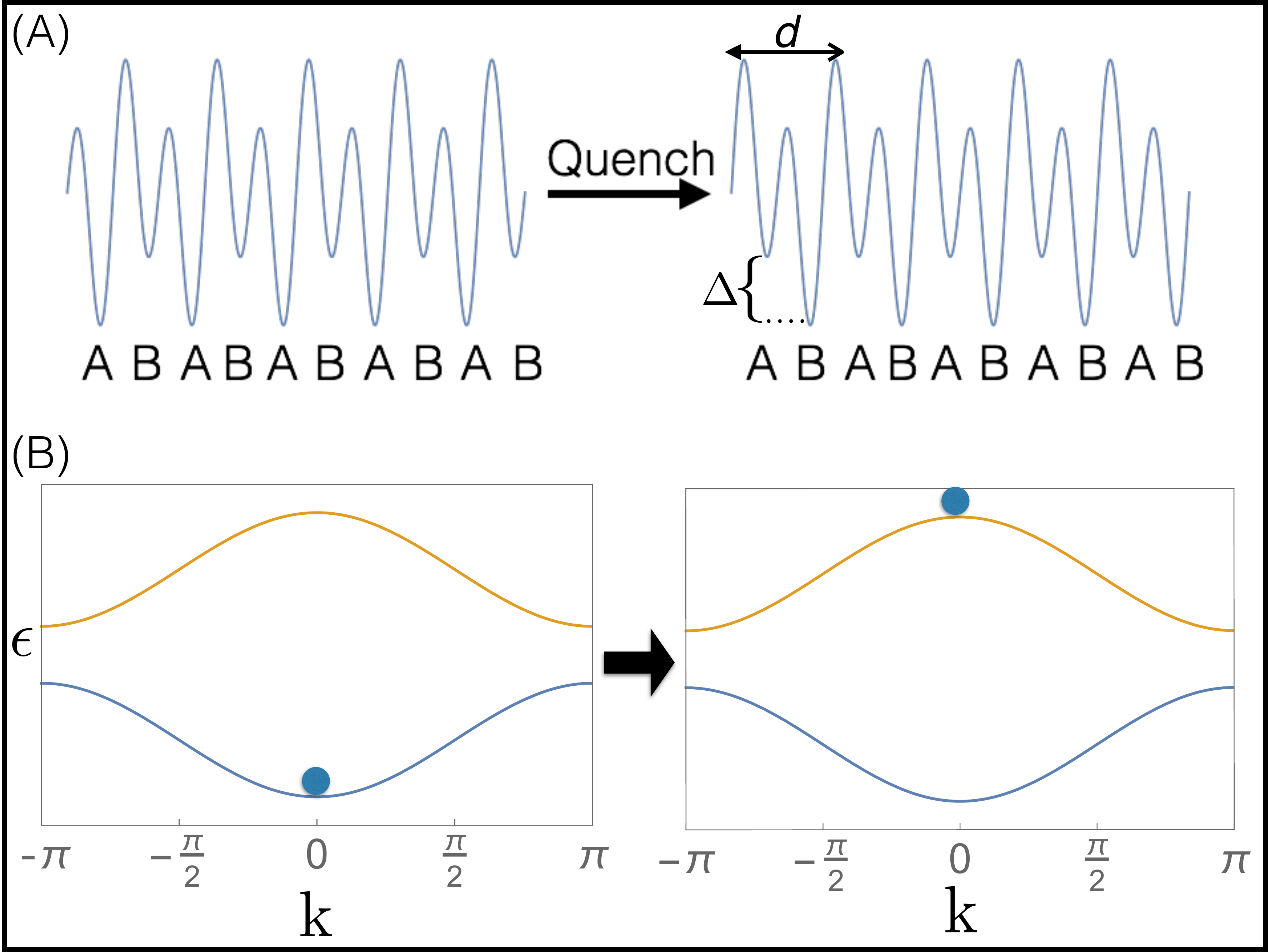}
\caption{(A) Envisioned setup for the experiment (B) Visualizing how the condensate gets transferred from the lower band to the highest energy state in the upper band after quench}
\label{setup}
\end{figure}

%\begin{figure}
%\includegraphics[scale=0.23]{fig1_1.pdf}
%\caption{(A) Quench Protocol: The Bose-Einstein condensate (BEC) is loaded in a bipartite lattice where the two sites of the unit cell are labelled A and B. Initially the A sites are much deeper than the B sites. By suddenly changing the lattice depths, the B sites are made much deeper than the A sites. This transfers the BEC to a higher band. (B) The condensate is initially in the lower band at {\bf k=0}. The quench protocol transfers it to the upper band at ${\bf k =0}$. The upper band BEC is initially unstable and it evaporates very fast. Eventually if the initial density of bosons is greater than a critical value, Bose-Einstein condensation happens at the band edge.}
%\label{setup}
%\end{figure}
Motivated by the MIT experiment \cite{KetterleHigherBand,KetterleHigherBand2}, and related experiments at Hamburg \cite{hemmerichNATP,HemmerichPRL2011,HemmerichPRL2015}, we model the dynamics of a BEC loaded into a double well optical lattice. A schematic of the setup is shown in Fig.\ref{setup}(A). 
 The single-particle Hamiltonian, $H_0$ describing this system is given by:
\bea\label{Hrealspace}
H_0 &=&  \int d^2 {\bf r_{\perp}} \sum_i \Delta(t) \,\, b_i^{\dagger}b_i -\left(J_1 a_i^{\dagger}b_i+J_2 a_i^{\dagger}b_{i-1}  + h.c.\right)\nonumber \\
&+& \frac{\hbar^2}{2m}\left(\nabla_{\perp} b_i^{\dagger}\nabla_{\perp} b_i + \nabla_{\perp} a_i^{\dagger}\nabla_{\perp} a_i\right)
\eea
where the lattice is in the z-direction. The transverse spatial components
are suppressed : $a_i = a_i({\bf r_{\perp}})$ 
% (b_i=b_i({\bf r_{\perp}}))$ 
is the annihilation operator for a boson at site $i$ of the $A$ %(B)
sub-lattice where $r_{\perp} = (x,y)$ and $\nabla_{\perp} = \hat{x}\partial_x + \hat{y}\partial_y$.  The operators $b_j$ have analogous meaning for the $B$ sublattice. 
For this paper, we consider the case $J_1=J_2=J^{\prime}$. \\

Before the start of the experiments, the energy offset between the A and B sites, $\Delta(t<0)=\Delta$ and the BEC is in the state ${\bf k =0}$ of the lowest band. The experimental protocol then involves  changing the lattice depths very fast such that after the quench, $\Delta(t>0)=-\Delta$.

The single particle Hamiltonian is diagonal in momentum space as shown in Appendix~\ref{kspace}, and the dispersion for the higher band is
\be\label{dis}
\epsilon_{\bf k} = 
%\sqrt{(\frac{\Delta}{2})^2 + (2 J^{\prime} \cos(k_z d/2)^2}- \frac{\Delta}{2} + \frac{\hbar^2 k_{\perp}^2}{2 m},
J(1+\cos(k_z d)) + \frac{\hbar^2 k_{\perp}^2}{2 m},
\ee
where $J = 2 (J^{\prime})^2/\Delta$,  and $d$ is the length of the unit cell.  Here
$\bf k_\perp$ can be arbitrary, but $-\pi/d<k_z<\pi/d$. The band eigenstates are also derived in Appendix~\ref{kspace}.

The $k=0$ state in the lowest band before the quench has nearly unit overlap with the post-quench $k=0$ state in the upper band, and  
 the quench projects the condensate into the higher band.
   A similar approach has been used to create an excited band BEC in a two-dimensional checkerboard lattice \cite{hemmerichNATP,HemmerichPRL2011,HemmerichPRL2015}.

As we argue below, the time-scale for atoms to equilibrate in the upper band is much smaller than band-relaxation.  Thus, we predominantly study single-band kinetics, using the dispersion in Eq.~(\ref{dis}).

\subsection{Interactions}
The kinetics are driven by  point interactions,
\begin{eqnarray}\label{pointinteraction}
H_{\rm int} &=& \frac{g}{2} \int d^3 r\, \psi^\dagger(r) \psi^\dagger(r) \psi(r)\psi(r),
\end{eqnarray}
where $g=4\pi\hbar^2 a_s/m$, with scattering length $a_s$.  The field operators, projected into our single band, are expressed as
\begin{equation}
\psi({\bf r}) = \sum_j \bar a_j({\bf r_\perp}) w(z-z_j)
\end{equation}
where $w(z)$ is the Wannier state and  $z_j=j d$ is the location of the $j$'th site. Neglecting the overlap between Wannier states on distinct sites, one
arrives at an effective delta-function interaction in each plane, which can be written as either an integral or a sum in momentum space:
\begin{eqnarray}\nonumber
H_{\rm int} &=& \frac{U}{2} \frac{V^2 d}{(2\pi)^9} \int \! d^3{ k_1} d^3 { k_2} d^3{ k_3} \,\, \bar a_{\bf k_1}^{\dagger}  \bar a_{\bf k_2}^{\dagger} \bar a_{\bf k_3} \bar a_{\bf k_1+k_2-k_3}\\ \label{hint}
&\sim& \frac{U}{2} \frac{d}{V} \sum_{k_1k_2k_3}  \bar a_{\bf k_1+k_2-k_3}^{\dagger} \bar a_{\bf k_3}^{\dagger} \bar a_{\bf k_2}\bar a_{\bf k_1}
\label{hk}
\end{eqnarray}
In the second form, the sum is over $k=2\pi n/L$ and $V=L^3$, where $L$ is a multiple of $d$.  The operator $a_{\bf k}$ is defined in Eq.~(\ref{fieldop}). 
%
%.
%Projecting the field operators into the higher band yields
%\begin{equation}
%\psi(r) \to  \sqrt{V}\int_{-\pi/d}^{\pi/d} \frac{dk_z}{2\pi} \int \frac{dk_x dk_y}{(2\pi)^2} e^{i {\bf k\cdot r}} a_{\bf k} W_{k_z}(z),
%\end{equation}
%where the Bloch state wavefunctions can be related to the Wannier states $w(z)$ via
%\begin{equation}
%W_{k_z}(z)=\sum_j w(z-z_j) e^{-i k_z (z-z_j)}
%\end{equation}
%and $z_j = d j$ are the positions of the $A$-sites.  In the limit that the Wannier states are small compared to the separation between sites,
%$W_{k_z}(z)  \approx w(z)$ becomes independent of $k_z$.  
%\begin{eqnarray}
%\frac{U_a}{A} \frac{V^3}{(2\pi)^3}\int \! d{\bf k_1} d {\bf k_2} d{\bf k_3} \,\, a_{\bf k_1}^{\dagger}  a_{\bf k_2}^{\dagger} a_{\bf k_3} a_{\bf k_1+k_2-k_3}
%\end{eqnarray}
In either case
\be
U = \frac{4 \pi \hbar^2 a_s}{m}\int dz \vert w(z) \vert^4 = \frac{4 \pi \hbar^2 a_s}{m d_a}. \nonumber
\ee
The last equality defines the characteristic width of the Wannier state $d_a$. Note, that in contrast to the standard Hubbard $U$, which is an energy, here $U$ has units of energy times length squared. This structure occurs because the atoms are free to move perpendicular to the lattice.
%The cross-sectional area of the cloud is $A=V/d$

%\begin{figure}
%\includegraphics[width=\columnwidth]{condplot50.jpg}
%\caption{
%{\color{red} May no longer be needed! (Try log plot on horizontal axis)}
%The figure shows the total number of bosons, $N$ and the number of bosons in the final condensate $N_{\pi}$ normalized by $N^{*}=0.87 JmV/\hbar^2d$. If $N/N^{*}$ is less than 1, then there is no final condensate.}
%\label{condcrit}
%\end{figure}

\section{Steady State} 

The long-time behavior in the upper band is solely determined by conservation laws.  After the quench, the kinetic energy is $E=2 N J$.  In the absence of band relaxation, 
the system will evolve so that there are $N_\pi$ atoms in the condensate at ${\bf k_c} = (k_x=k_y=0,k_z=\pi/d)$ and $N_{\rm nc}$ non-condensed atoms.  According to the higher band dispersion, only the non-condensed atoms contribute to the kinetic energy.  Neglecting interactions, their number and kinetic energy are
\begin{eqnarray}
\label{eqnscond}
\frac{N_{\rm nc}}{V} &=&  \int\!\! \frac{d^3 {k}}{(2 \pi)^3} f_k
%1/\left(\exp(\beta(\epsilon_{\bf k} - \mu)-1\right) \nonumber\\
= \frac{\rho_0 J}{4\pi^2} F(\beta J)
=  %\frac{\eta}{V} %\frac{ J m V}{(2\pi)^2 \hbar^2 d} 
\int\!\! d\epsilon \,\rho(\epsilon) f(\epsilon)
%\nonumber
\\
\frac{E_{\rm nc}}{V} &=&  \int\!\! \frac{d^3 k}{(2 \pi)^3} \epsilon_{\bf k} f_k
%/\left(\exp(\beta(\epsilon_{\bf k} - \mu)-1\right) \nonumber\\
%&=& J\pi \frac{2m V}{\hbar^2 \beta d} G(\beta J) \nonumber\\
=\frac{\rho_0 J^2}{4\pi^2}
G(\beta J)
= % \frac{J \eta}{V} %\frac{J m V}{(2\pi)^2 \hbar^2 d} 
 \int d\epsilon \,\epsilon \rho(\epsilon) f(\epsilon), \nonumber
\end{eqnarray}
where %$\eta=JmV/(4\pi^2\hbar^2 d)$, and 
the density of states is
\begin{equation}\label{dos}
\tilde\rho(\epsilon)=\frac{\rho(\epsilon)}{\rho_0}=
  \left\{ \begin{array}{lr}1&\epsilon\geq 2 J\\
\frac{1}{\pi}\cos^{-1}\left(1-\frac{\epsilon}{J}\right)&\epsilon<2J
\end{array}  \right.
\end{equation}
with $\rho_0=m/\hbar^2d$. The characteristic length of the system is $(\rho_0 J)^{-1/3}$.

Equations~(\ref{eqnscond}) 
define the dimensionless functions $F$ and $G$.  The Bose occupation factors are
 $f_k=f(\epsilon_{\bf k})=(\exp(\beta \epsilon_{\bf k} )-1)^{-1}$, in which we have taken the chemical potential to vanish, corresponding to the conditions for having
 a condensate at $\bf k_c$.  
 The density of states is three-dimensional at small $\epsilon$, $\rho(\epsilon\to0)\sim \sqrt{\epsilon}$, and two-dimensional at large $\epsilon$,  $\rho(\epsilon\to\infty)\sim \epsilon^0$.
 
 The functions $F$ and $G$ are readily evaluated numerically.  
 %In particular, straightforward manipulations
 %convert them into the compact one-dimensional integrals
 %\begin{eqnarray}\label{feq}
 %F(x)&=& -\frac{1}{x} \int_{-\pi}^\pi \frac{ds}{4\pi^2}\, \ln\left(1-e^{-x (1+\cos s)}\right)\\\label{geq}
 %G(x)&=&\frac{1}{x^2} \int_{-\pi}^\pi \frac{ds}{4\pi^2}\, \left[
%{\rm Li}_{2}\left(e^{-x (1+\cos s)}\right) \right.\\\nonumber&&\quad\left.-x(1+\cos s)  \ln\left(1-e^{-x (1+\cos s)}\right)
 %\right],
 %\end{eqnarray}
 % where ${\rm Li}_2(s)=\sum_{j=1}^\infty s^j/j^2$ is the dilogarithm function.
 % 
%The magnitude of the transverse momentum is $\hbar k_{\perp}$, $V$ is the volume of the  system.  
The final inverse temperature $\beta$ is found by solving $E=2J (N_{\rm nc}+N_\pi)$, or $N_\pi=\frac{J\rho_0 V}{4\pi^2} (G(\beta J)/2- F(\beta J))$.   We find $N_\pi>0$ if and only if $\beta J<0.35$.  This corresponds to $N>N^*$, where
\begin{equation}\label{nstar}
N^*=0.85 \frac{J m V}{\hbar^2 d}= 0.85 \rho_0 J V.
\end{equation}
We conclude that if the initial number of bosons is greater than $N^{*}$, then the final state has a condensate, while if the initial number of bosons is smaller than $N^{*}$, then the final state does not have a condensate. 

As one would expect, the threshold $N^*$ is extensive.  The condition $N=N^*$ can be understood by noting that the average transverse kinetic energy after relaxation is of order $J$, corresponding to a DeBroglie wavelength of order $\lambda=\hbar/\sqrt{2 m J}$.  The threshold for condensation corresponds to when the average separation between particles in each 2D pancake is comparable to $\lambda$.

Interactions will somewhat move the threshold, but should not change the general behavior.

\begin{figure}[tb]
\includegraphics[width=\columnwidth]{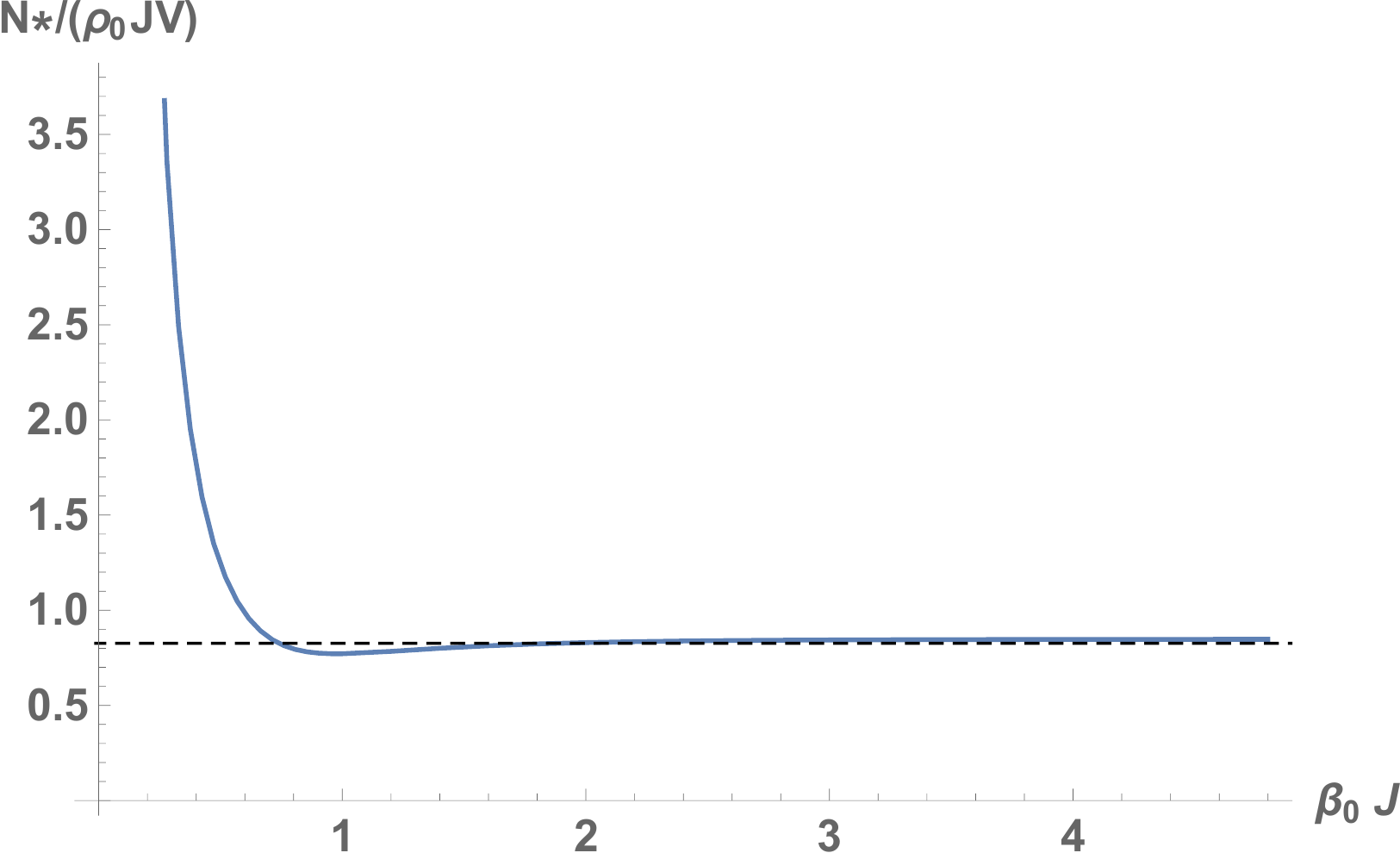}
\caption{Threshold $N^*$ as a function of initial state temperature $\beta_0 J$. The dashed line shows $N^*$ for $\beta_0 J \to \infty$}
\label{finitetempfig}
\end{figure}

In the limit $N\gg N^*$, the fraction of non-condensed atoms becomes small.  In that limit one can expand Eqs.~(\ref{eqnscond})  in powers of $x$ (or $\beta J$): $F(x)\to -(2\pi/x)\ln(x)$ and $G(x)\to 2\pi/x^2$. Thus in this limit, the final temperature  becomes very large compared to $J$:
$\beta J \to \sqrt{\rho_0 V J/(4\pi N)}$ The noncondensed fraction scales as
$N_{\rm nc}/N \to N^{-1/2}\log N$ as $N\to\infty$.

\subsection{Finite Temperature}
Our argument can be readily modified to account for thermal effects in the initial state.
%Finite temperature of the initial state of condensed particles in the lower band would also move the threshold $N^*$ required to form a condensate in the higher band.

Given an initial temperature $\beta_0$, the initial distribution of particles will be given by 
\begin{equation}
    f_{\epsilon_k^L}(\beta_0)=\frac{1}{e^{\beta_0 (\hbar^2 k_{\perp}^2/2m -J(1+cos(k_z d)))}-1}.
\end{equation}
Since the occupations are based on the lower-band energies, this is very different than the equilibrium occupation of the higher band.

Given these occupations, the condensate number and the total energy after the quench are
\begin{eqnarray} \label{steady1}
    N_0 &=& N - \int \frac{d^3 k}{(2\pi)^3} f_{\epsilon^{L}_k} (\beta_0)\\
\label{steady2}
    E &=& 2JN_0 + \int \frac{d^3 k}{(2\pi)^3} \epsilon_k f_{\epsilon^{L}_k}(\beta_0).
\end{eqnarray}
As before, both number and energy are conserved in the dynamics, and the $\beta J$ describing the equilibrium distribution is found from Eq.~(\ref{eqnscond}), setting $N = N_{nc} + N_{\pi}$ and $E = E_{nc}$.  In particular, $N^*$ the threshold number of particles to find a condensate, is produced by setting $N_\pi=0$.

%In the higher band, the equilibrium temperature $\beta J$ for a given $\beta_0 J$ would be determined by setting $N = N_{nc} + N_{\pi}$ and $E = E_{nc}$ in Eq. \ref{steady1} and \ref{steady2} where $N_{nc}$ and $E_{nc}$ are same as in Eq. \ref{eqnscond} and solving for $N_{\pi} \geq 0$.

%Here, $\epsilon^{L}_k$ is the energy in the lower band which is explicitly given in Appendix \ref{kspace} 

Figure \ref{finitetempfig} shows how $N^*$ varies with the initial temperature $\beta_0 J$.  The result is non-monotonic.  When $k_B T_0<J,$ increasing the initial state temperature before quenching into the higher band actually reduces the total energy:  Upper band atoms with $k_z>0$ have smaller energy than those with $k_z=0$.  This results in a smaller $N^*$.

Once $k_B T_0>J$, further increasing the initial temperature in the lower band results in a higher upper-band energy:  The relevant excitations are transverse to the lattice. This results in a larger $N^*$

Clearly if $k_B T_0<J$, finite temperature effects are small, and it is reasonable to neglect them.

\section{Higher Band Kinetics} 
\begin{figure*}
\includegraphics[height=12cm]{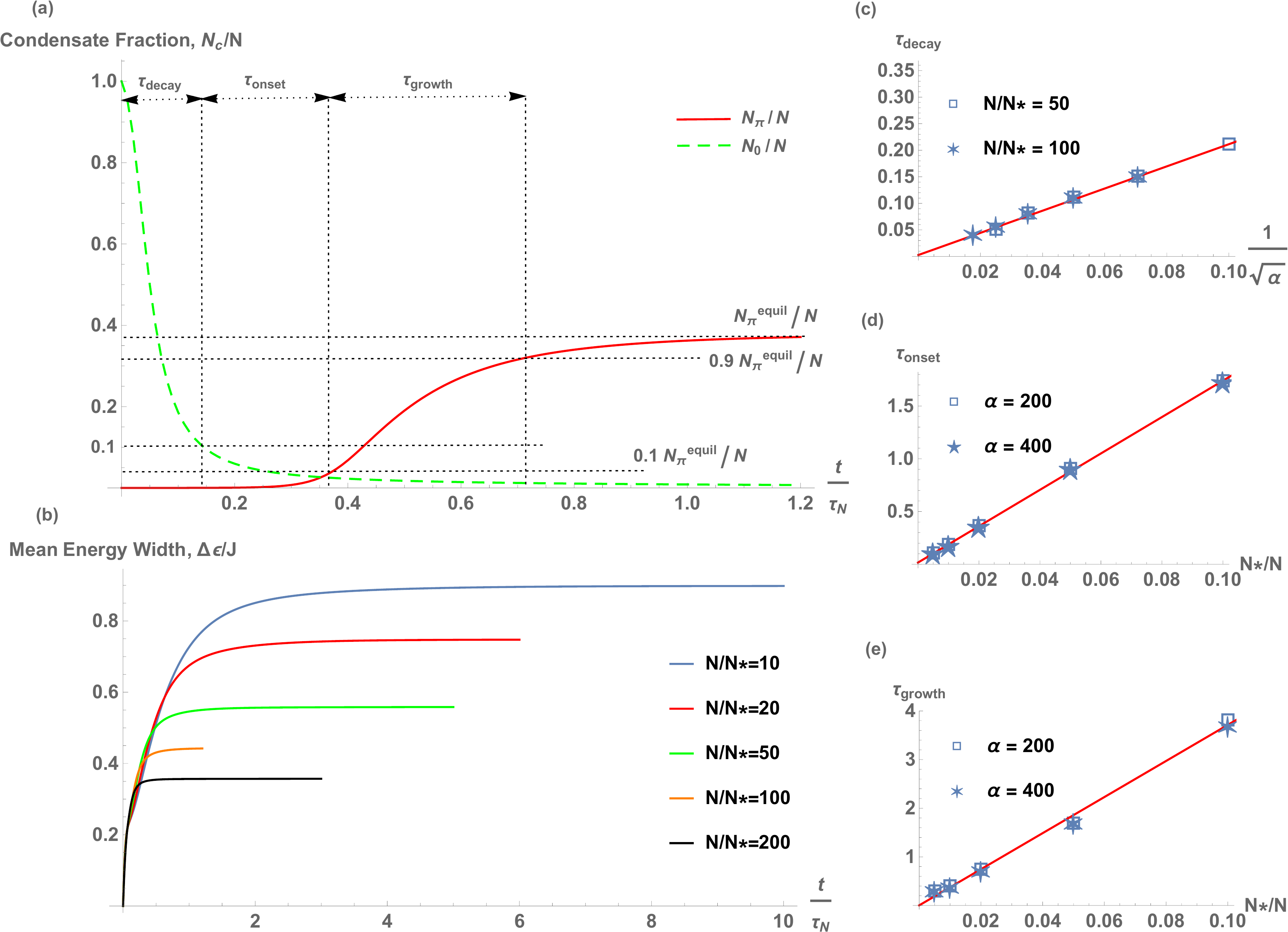}
\caption{(color online) (a) Fraction of particles in $k_z = 0$ condensate (green,dashed) and $k_z=\pi$ condensate (red,solid) plotted against dimensionless time, $t/\tau_N$ for $N/N^* = 100$ and $\alpha = 200$. Three different timescales can be seen: $\tau_{decay}$, the decay of $k_z=0$ condensate, $\tau_{onset}$, onset of formation of $k_z=\pi/d$ condensate; and $\tau_{growth}$, the growth of $k_z=\pi/d$ condensate (b) Mean energy width, $\Delta \epsilon/J$ of the distribution functions, $f(\varepsilon)$ versus $t/\tau_N$ for different $N/N^*$ values with $\alpha=200$. $N/N^*$ increases from top to bottom. (c) $\tau_{decay}/\tau_N$ (d) $\tau_{onset}/\tau_N$ (e) $\tau_{growth}/\tau_N$}
\label{mainfig}
\end{figure*}

%\subsection{Boltzmann Equation}
%Having established the condition for the existence of a higher band condensate at long times, we proceed to model the kinetics. 
%Here we derive the relevant kinetic equations, which we subsequently analyze and numerically integrate.
%There are three relevant kinetic processes for modeling the relaxation dynamics in the upper band: $\Gamma_0$, the rate of scattering out of the ${\bf k = 0}$ condensate; $\Gamma_{\rm in}$,  the rate of scattering into the ${\bf k = k_c}$ condensate; and $\Gamma_{\rm th}$, the rate at which the non-condensed atoms thermalize.  For typical experimental parameters, we find $\Gamma_0>\Gamma_{\rm in}> \Gamma_{\rm th}$, though one can also reasonable find parameters where $\Gamma_0>\Gamma_{\rm th}> \Gamma_{\rm in}$.  In this section we derive expressions for these rates, and discuss the interplay of these processes.

%The initial mean-field energy of the condensate is $E_{\rm mf}=U N^2 d/2V$.   For recent experiments, such as the one in \cite{hemmerichNATP}, one has  $d_a\sim 100$nm, $a_s \sim 5$nm, $d\sim 500$nm, and $N/V = 10^{19} \, {\rm m^{-3}}$, and hence $E_{\rm mf} \sim 0.2 E_{\rm kin}$, where $E_{\rm kin}=4J N$ is the average kinetic energy per particle.  For these parameters mean-field effects are relatively unimportant, and can be ignored.  
%In section~\ref{strongint} we will discuss the case of tighter optical lattice (such as in \cite{KetterleHigherBand})  where mean-field effects dominate.
%We use these parameters for all the calculations in this paper.\\

Neglecting coherences between different 
momenta, one can use Fermi's golden rule to derive a quantum Boltzmann equation \cite{kadanoff}.
It is a standard practice to make an ergodic approximation \cite{ergodic1, ergodic2, ergodic3} where all states of the same energy are taken to be equally occupied. This approximation postulates that equilibration between modes of same energy is fast compared to energy redistribution. 
%Typically equilibration between modes with the same energy is commonly postulated to be fast compared to energy redistribution, motivating a standard ergodic approximation, where all states of the same energy are taken to be equally occupied.
We define $f(\varepsilon)$ (or $f_{\varepsilon}$) to be the occupation of modes with energy $\epsilon=J\varepsilon$.  We separate out the mode with $k=0$, defining $M=N_0/N$ to be the fraction of particles in that condensate.  In Appendix~\ref{BTE},we show,
\begin{widetext}
\begin{eqnarray}\label{mastereq}
\frac{\partial f(\varepsilon_1)}{\partial \tilde t}
&=&\frac{1}{\tilde \rho(\varepsilon_1)}
\int \frac{d\varepsilon_2d\varepsilon_3d\varepsilon_4}{(2\pi)^3}\Pi^{12}_{34\,}2\pi\delta(\varepsilon_{1}+\varepsilon_2-\varepsilon_{3}-\varepsilon_4)\,\left[
f_3 f_4 (1+f_1)(1+f_2)-f_1f_2(1+f_3)(1+f_4)
\right]\nonumber\\\label{kinetic}
&&+0.85\frac{N}{N^*}M^2 (1+2 f_1)\frac{\bar\Gamma}{(\varepsilon_1-2)^2+(\bar \Gamma/2)^2}.
\end{eqnarray}
\end{widetext}
The second line corresponds to processes where two particles scatter out of the $k=0$ condensate, while the first line includes processes where particles with energy $\varepsilon_1$ and $\varepsilon_2$ scatter into $\varepsilon_3$ and $\varepsilon_4$, or vice versa.  Throughout, $f_j=f(\varepsilon_j)$, and the dimensionless density of state $\tilde \rho$ is defined in Eq.~(\ref{dos}).
\begin{equation}
    \int d\varepsilon \tilde\rho(\varepsilon)f(\varepsilon) = N_{nc}/(\rho_0 J V) = 0.85 N_{nc}/N^*
\end{equation}

The rate of scattering out of the $k=0$ condensate is parameterized by
\begin{eqnarray}\label{gamma}
\bar \Gamma &=& -0.85\frac{N/N^*}{\alpha} \frac{1}{M}\frac{dM}{d \tilde t} \label{nounitdecay2}\\\nonumber
&=& 0.85\frac{N/N^*}{\alpha} M \int d\varepsilon \tilde\rho(\varepsilon) (1+2f(\varepsilon))\frac{\bar \Gamma}{(\varepsilon-2)^2 + (\bar \Gamma/2)^2}.  \end{eqnarray}
The Lorentzians in Eq.~(\ref{kinetic}) and (\ref{nounitdecay2}) accounts for broadening due to the short condensate lifetime.

The dimensionless coefficient $\Pi^{12}_{34}=\Pi(\varepsilon_1,\varepsilon_2,\varepsilon_3,\varepsilon_4)$ is derived in Appendix~\ref{MatrixElement}.  Aside from a multiplicative factor of  $N^*/N$, it only depends on the scaled energies, and no other parameters.  When all scaled energies are smaller than $2$, it reduces to a standard 3D result \cite{landau}, 
\begin{equation}
\Pi_{\varepsilon_{\rm max}<2}\propto \sqrt{\varepsilon_{\rm min}},
\end{equation}
where $\varepsilon_{\rm min}$ and $\varepsilon_{\rm max}$ are the smallest and largest of the $\varepsilon_j$.  For large energies it becomes an elliptic function. We use an approximate form (explicitly given in the appendix) which interpolates between these two expressions.

Times have been scaled, $\tilde t=t/\tau_N$, where
\begin{equation}\label{timescale}
\tau_N= \frac{2\hbar V} {N \rho_0 (U d)^2}= \frac{2}{(4\pi)^2}\frac{d_a}{n d}  \frac{1}{a_s^2} \frac{m d_a}{\hbar}.
\end{equation}
This scale can be interpreted as a microscopic collision time, $\tau_N\sim 1/(n_{\rm eff} \sigma v)$, where $n_{\rm eff}=n d/d_a$ is the effective density.  The enhancement factor $d_a/d$ reflects the fact that the Wannier states are compressed in one direction.  The cross-section $\sigma=4\pi a_s^2$ is proportional to the square of the scattering length.  In this interpretation the characteristic velocity is proportional to $\hbar/m d_a$.  There are other possible velocities in the problem, and {\em a priori} it  is not 
 obvious which one to use.  Nonetheless Eq.~(\ref{timescale}) is a scaling which simplifies the equations.
 
 In addition to $N/N^*$, there is only one other dimensionless parameter in these equations, 
  \begin{equation}
     \alpha=\frac{N \tau_N}{\hbar V \rho_0}= 0.85 \frac{J \tau_N}{\hbar}\frac{N}{N^*} =\frac{2}{(4\pi)^2} \left(\frac{d_a}{a_s}\right)^2.
 \end{equation}
 The last expression is most transparent:
 recall, $a_s$ is the scattering length, and $d_a$ is the width of the Wannier states.  Typically, $\alpha\sim 10$, though it can readily be increased or decreased by an order of magnitude by changing the lattice depth or employing a Feshbach resonance.
 In a given experiment, $\frac{N}{N^*}$ is varied by changing the number of atoms, or the lattice depth -- see Eq.~(\ref{nstar}).
 
 Our derivation breaks down if the condensate lifetime becomes significantly smaller than $\hbar/J$.  
   In section~\ref{decay}, we analyze the decay process, and find $\tau_{\rm decay}\sim \tau_n/\sqrt{\alpha}$. Consequently, we require that $\alpha$ is not too small compared to $(N/N^*)^2$.  Accurately modeling the small $\alpha$ limit would require keeping track of the coherences between the modes occupied during the evaporation process.  Nonetheless, we expect our results to capture much of the physics, even in that limit.

 \section{Results}
 We numerically integrate Eq.~(\ref{mastereq}). The algorithmic details for this are in Appendix~(\ref{discretization}).  Fig.~\ref{mainfig} (a),\ref{mainfig}(b) show typical time-series for the $k=0$ condensate fraction, the $k=\pi/d$ condensate fraction, and the width of the energy distribution $\Delta\varepsilon = \frac{N^*}{N} \sqrt{\int d\varepsilon \rho(\varepsilon)f(\varepsilon)(\varepsilon-2)^2}$.  
  Four separate timescales are apparent:
 $\tau_{\rm decay}$ is the timescale for decay of the $k=0$ condensate; $\tau_{\rm onset}$ is the characteristic time for the $k=\pi/d$ condensate to start growing, $\tau_{\rm growth}$ is the timescale for the $k=\pi/d$ condensate to grow to its equilibrium value, and $\tau_E=J/(d (\Delta\varepsilon)/dt)$ is the inverse slope of the energy-width curve.
 
 Numerically we find that $\tau_{\rm decay}\sim \tau_N/\sqrt{\alpha}$, and $\tau_{\rm onset}\sim \tau_{\rm growth} \sim \tau_N N^*/N$, and $\tau_E\sim \tau_N$ (see Fig.~\ref{mainfig} (b,c,d,e). Thus when $\alpha\gg 1$ and $N > N^*$ there is a clear separation of scales.  In sections~\ref{decay},\ref{redistribute} and \ref{growth}, we give analytic arguments for the scaling of the decay and growth processes.

 \subsection{Decay}\label{decay}
 
 The first stage of the dynamics, as illustrated in Fig.~\ref{mainfig}(a) is the decay of the $k=0$ condensate.  There, pairs of particles scatter to states whose energies are near $2J$. 
 
 To understand the scaling of this process as shown in Fig.~\ref{mainfig}(c), we neglect the first line of Eq.~(\ref{kinetic}):  As is verified by the numerics, the redistribution of energy amongst the non-condensed particles is slow compared to the evaporation.   Throughout this initial stage, the function $f(\varepsilon)$ will be peaked about $\varepsilon=2$, with height $f_2$ and width of order $\bar\Gamma$.  Number conservation, Eq.~(\ref{conservation}), implies that $f_2\sim \frac{N}{N^*} (1-M)/\bar\Gamma$, where $M=N_0/N$ is the $k=0$ condensate fraction.  Recall that our arguments apply when $\alpha$ is large, and hence the rate $1/\tau_{\rm evap}= J \bar\Gamma/(\hbar)$ will be small.  Thus $f_2$ will be large compared to 1, and in Eq.~(\ref{kinetic}) we can replace $1+2f\approx 2f$.  The integrand in Eq.~(\ref{nounitdecay2}) will have height $f_2/\bar\Gamma$, and width $\bar\Gamma$, and hence the integral is of order $f_2$.  Thus one expects $\bar\Gamma \sim \frac{N/N^*}{\sqrt{\alpha}} \sqrt{M(1-M)}$, as long as $M$ is not too close to 1.  The characteristic time-scale for decay of the $k=0$ condensate is found by taking $M(1-M)$ to be of order 1, which yields $\tau_{\rm evap}=\hbar/(J\bar\Gamma)\sim\tau_N/\sqrt{\alpha}.$ 
 
 \subsection{Energy Redistribution}\label{redistribute}
 The second stage of the dynamics, as seen in Fig.~\ref{mainfig}(b), is the redistribution of energy amongst the non-condensed particles.  At short and intermediate times, the energy-width of the distribution function grows roughly linearly in time.  The slope of this curve is of order $J/\tau_N$, consistent with the fact that the typical energy is $2J$ and the characteristic scattering time is $\tau_N$.  The energy-width saturates at long time.  The time-scale for saturation is roughly the onset time for growth of the $k=\pi/d$ condensate.
 
 \subsection{Growth}\label{growth}
 
%In section ??, we explore these full sets of coupled equations.  As previously explained, our Boltzmann equation approach requires that the condensate lifetime is large compared to the inverse bandwidth.  Given that the characteristic time is $\tau_N$, we therefore require
 %$tau_N \gg 1/J,$ which implies that $\alpha \gg 1$. In that limit we will find a separation of timescales, where the condensate decays on a timescale of order $\tau_N/\sqrt{\alpha}$, while the system equilibrates on a timescale of order $\tau_N$.  The time-scale for onset of the recondensation process and the growth of the final condensate scales as $\tau_N/(N/N^*)$.

 The scaling of the onset and growth times as seen in Fig.~\ref{mainfig} (d),\ref{mainfig}(e) both are consequences of Bose stimulation.  Once the $k=0$ condensate evaporates, the non-condensed particles redistibute their energies.  A microscopic seed forms at $k=\pi/d$ in a time of order $\tau_N$.  The number of particles in that seed will scale linearly with the density, and will therefore be proportional to $N$.  This seed then grows exponentially, and the time that it takes to become macroscopic will be inversely proportional to the initial number.  Hence $\tau_{\rm onset}\sim \tau_N N^*/N$.  The timescale for growth will also scale in this manner.

\section{Decay to the lower band} \label{bandrelaxation}

Our analysis is predicated on the dynamics within the band being fast compared to the inter-band decay.  Here we estimate that decay rate, finding that the ratio of the inter-band and intra-band rates is proportional to $(J^\prime/\Delta)^2$. Since in the experiments $(J^\prime/\Delta)\ll 1$ \cite{KetterleHigherBand2}, there is a large separation of scales.

This suppression comes from the poor spatial overlap between the upper band wavefunctions (which are predominantly on the A sublattice) and the lower band wavefunctions (predominantly on B).   

%The higher band BEC is unstable and all the bosons in the upper band will eventually relax to the lower band \cite{PaulHigherBand}. 

The loss of atoms from the condensate in the upper band at ${\bf k=\pi}$ to the lower band is driven by the interaction term \cite{PaulHigherBand}, and the rate can be calculated using Fermi's golden rule.  The leading process involves two upper band $k=\pi$ atoms scattering to produce a lower band atom with momentum $k$, and an upper band atom with momentum $k^\prime$.  Using the dispersion calculated in Appendix~(\ref{kspace}), the energy of this final state only depends on the transverse momentum, $\epsilon_f = -\Delta+2\hbar^2 k_\perp^2/2m$.  The matrix element is calculated by substituting the operators for the Bloch states from Appendix~\ref{kspace} into the interaction Hamiltonian.  Taking $N_\pi\approx N \gg 1$ and assuming that none of the  lower band states are macroscopically occupied, we can repeat the argument in  Appendix~\ref{evap1}
that we used to calculate intra-band decays, and find,
%. To see this, we can refer to the point interaction term in Eq.~(\ref{pointinteraction}) where two particles from the higher band condensate scatter into two particles in the lower band where the higher and lower band states are given in the Appendix (\ref{kspace}) in Eq.~(\ref{hb})
 %and Eq.~(\ref{lb}). The leading interaction term in momentum space gives:
%\bea
%H_{\rm int}
% &\approx&  \frac{U_a}{V} \frac{4 J'^2}{\Delta_0^2}\int \! d{\bf k_1} d {\bf k_2} d{\bf k_3} \,\, \overline{b}_{\bf k_1}^{\dagger}  \overline{b}_{\bf k_2}^{\dagger} \overline{a}_{\bf k_3} \overline{a}_{\bf k_1+k_2-k_3} \nonumber \\
%\eea
%Thus, the loss of atoms from the condensate at ${\bf k = \pi}$ is given by
\begin{eqnarray}
\Gamma_{ab} &&= \frac{-1}{N_\pi} \frac{d N_\pi}{dt} = \frac{2 \pi}{\hbar} \sum_f |\langle \psi_f \vert H_{{\rm int}} \vert \psi_i \rangle|^2 \delta(\epsilon_f-\epsilon_i). \nonumber\\
=&& \frac{N}{V \hbar}\int\!\!\frac{d^3 k}{(2\pi)^2}
\left(
\frac{2 J^\prime U_a \cos{(k_z d/2)}}{\Delta}
\right)^2 \delta(\frac{\hbar^2k_\perp^2}{m}-\Delta).\nonumber \\
=&& \left(\frac{2 J^\prime}{\Delta}
\right)^2 \frac{N m d U_a^2}{2 V \hbar^3} 
= \left(\frac{2 J^\prime}{\Delta}
\right)^2 \frac{1}{\tau_N}
\label{fincond}
\end{eqnarray}
%The initial state differs from the final state by having two particles scattered out of the $k=\pi$ condensate  to lower band states with momentum $k$ and $-k$.  The lower band dispersion is $\epsilon_k^\ell= -\Delta_0-J(1+\cos(k_z d)) +\hbar^2 k_\perp^2/2m$.  The factor in parenthesis is the matrix element, found by writing the field operators in Eq.~?? in terms of the operators for the two bands in Eqs.~??.  Intuitively, the states in the upper and lower bands have atoms occupying either the high-energy or low-energy sites of the dimerized lattice.  The amplitude for finding one on the other sublattice is of order $J^\prime/\Delta_0$.  Two particles are involv
%We obtain:
%\be
%\Gamma_{ab} = (\frac{2 J'}{\Delta_0})^4\frac{2 m V}{\hbar^3 d}U_{a}^2 N_{\pi} 
%\ee
As already explained, the factor
%In comparison to our typical decay rate within the higher band, this contains an additional square power of 
$({2 J'}/{\Delta})^2$ is typically much much smaller than 1, implying that the decay from the higher band ($\tau_{ab} \sim 1/\Gamma_{ab})$ is slow compared to the kinetics within the higher band.

%For typical experimental parameters \cite{KetterleHigherBand,KetterleHigherBand2}, $(\frac{2 J}{\Delta}) \sim 10^{-3}$ and for $\frac{N_{\pi}}{N} \sim 0.5$, we get $\Gamma_{ab} \sim 4 s^{-1}$. This timescale is much slower than all the relevant timescales for the upper band kinetics.

\section{Experimental Signature}

\begin{figure*}[t]
\includegraphics[height=8cm]{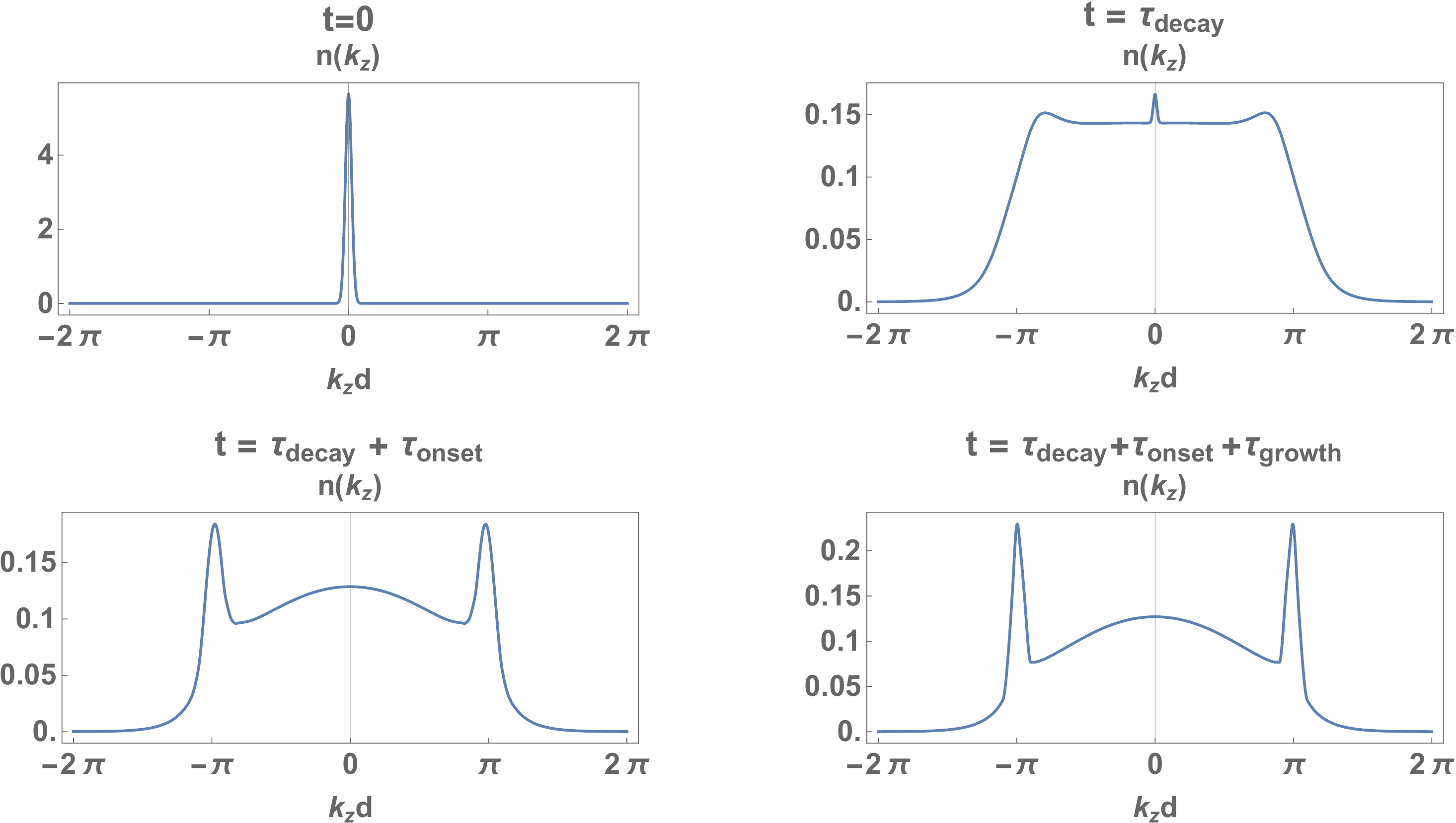}
\caption{Simulated time-of-flight images showing momentum space density of atoms along $k_z$ for $N/N^*=100$ and $\alpha=200$. (Area under the curves has been normalized to 1 in the figure)
The t=0 image represents time just after the quench and then the time-of-flight expansion is shown for elapsed times, $t=\tau_{decay}$, $t=\tau_{decay}+\tau_{onset}$ and $t=\tau_{decay}+\tau_{onset}+\tau_{growth}$, corresponding to the dotted vertical lines in Fig.~\ref{mainfig}(a)}
\label{tof}
\end{figure*}

A direct way to verify these kinetics is to experimentally measure the time-dependent momentum distribution through time-of-flight (TOF) expansion.  After free expansion for time $\tau$ one measures the column-density of atoms, 
\begin{equation}
n^c_{{\rm TOF}}(z) = \int \!\!dxdy\, n_{{\rm TOF}}({\bf r}).
\end{equation}
 Defining $ k_z=m z/(\hbar \tau)$, the column density is related to the {\em in-situ} momentum density of the trapped atoms at the time of release \cite{tofinsitu}, $t$,
\begin{eqnarray}
n_{{\rm TOF}} &\propto&
|w(k_z)|^2 
\int\!\!d k_x d k_y\,
n_{{\rm trap}}({\bf k},t)
\label{tofe}\\
&=& |w(k_z)|^2 
\int_{J(1+\cos{k_z d})}^{\infty}\!\!d\epsilon f(\epsilon,t)\\
\nonumber &&+ |w(0)|^2 N_0(t) \delta(k_z)
\end{eqnarray}
Here $w(k)$ is the Fourier transform of the Wannier function in the lattice, and when it is an argument of the distribution function, $k_z$ is projected into the Brillioun zone.

We numerically integrate the distribution functions calculated from Eq.~(\ref{mastereq}).  Figure~\ref{tof} shows the expected time-of-flight images at different times.  During the evaporation phase, the image is dominated by a delta-function peak at $k_z=0$.  In an experiment this peak has a non-zero width, set by the finite system size and the finite expansion time.  In Fig.~\ref{tof}, we use a Gaussian of width $0.01 k_z d$.  As the condensate evaporates, a halo representing the non-condensed particles appears.  As the particles redistribute themselves, structures form, and well before a $k_z=\pi/d$ condensate appears, one sees peaks near $k_z=\pi/d$.  These peaks sharpen over time as phase coherence develops on longer length scales.  A true condensate at $k_z=\pi/d$ would be characterized by delta-function peaks.  Again, finite system size and expansion time would spread out these delta-functions.  In our numerics the sharpness is limited by the resolution of our discretization of the energy.

For the plots in Fig.~\ref{tof}, we use a Gaussian Wannier state corresponding to a lattice depth of $5 E_R$ where $E_R$ is the recoil energy. We choose $N/N^*=100$ and $\alpha=200$

\section{Beyond Upper Bands}\label{upperbandsec}
As already discussed, our model can be used in settings other than the upper band of an optical lattice.  One example is the experiments of Lignier et al from Pisa~\cite{Pisa} where an optical lattice loaded with a condensate is sinusoidally shaken to dynamically change the tunneling amplitude J between nearest neighboring sites. They can flip the sign of the effective tunneling amplitude and invert the band. This situation is similar to our model where the condensate is promoted to the higher band.\\

In that experiment, the optical lattice has a spacing of 426 nm and is loaded with Rb-87 atoms. The lattice depth is about $9E_R$ where $E_R$ is the recoil energy of the lattice. The number density is of the order $10^{14} cm^{-3}$, which is typical of cold gas condensates and the s-wave scattering length between the atoms is about 5 nm. The timescale $\tau_N$ that we get from these parameters would be about 10 ms. While the published data does not include a detailed study of timescales, our estimate is consistent with the observation of a double peak structure in time of flight after a few ms.  Other single band realizations are likely to have similar parameters.

%The plot shows the momentum space density in the z direction at different times for $\frac{N}{N^*} = 100$. 

%Initially, everything is at $k_z=0$. At $t = 0.1\tau_N$, most of the condensate has decayed with other momentum states getting occupied. At $t = 0.4\tau_N$, we can already see the state $k_z=\pi$ getting occupied. But that is because it is the lowest energy state. At time $t = \tau_N$, the system has equilibrated and there is macroscopic occupation of the $k_z=\pi$ state, which would produce a sharp peak at $k_z=\pi$. 

%The time dependent momentum space density would be similar for other values of $\frac{N}{N^*}$. There would be a peak at $k_z=\pi$ at equilibrium but only when there is macroscopic occupation would there be a delta function peak here. 

\section{ Summary} 
We modelled the dynamics of a non-equilibrium condensate formed in the highest energy state of an excited band in an optical lattice. %These kinetics are important for producing exotic states of matter.  
We find that there is a critical particle number, below which the final state has no condensate.  We  derive kinetic equations and use them to calculate the time-dynamics of this system. We find  three distinct  timescales: a fast timescale over which the initial condensate evaporates, an intermediate timescale over which collisions occur, and slower timescale over which a new condensate grows. This scenario is very different from more conventional paradigms of order parameter dynamics, for example involving an order parameter ``rolling down" a potential hill \cite{chaikinlubensky} or evolving through a modulational instability\cite{ModInsRev,KevrekedisModIns}. This kinetic path is likely important in other experiments such as those involving shaken lattices\cite{ChinScience2016,ChinFloquetNew,Chin2,Chin3} or soliton formation\cite{HuletModIns}. 

We show how these processes can be seen in time-of-flight expansion images, allowing a direct experimental verification of our predictions.

\section*{Acknowledgements}  This work was supported by the NSF Grant PHY-1806357 and the ARO-MURI Non-equilibrium Many-body Dynamics Grant W9111NF-14-1-0003.

\appendix

\section{Derivation of Operators for Bloch Eigenstates and Band Dispersions}\label{kspace}
Here we explicitly give the momentum-space representation of the single particle Hamiltonian in Eq.~(\ref{Hrealspace}). 

%Before the start of the experiments, the energy offset between the A and B sites, $\Delta(t<0)=\Delta$ and the BEC is in the state ${\bf k =0}$ of the lowest band. The experimental protocol then involves  changing the lattice depths very fast such that after the quench, $\Delta(t>0)=-\Delta$. 

The real-space Hamiltonian is given in Eq.~(\ref{Hrealspace}).  We define momentum space field operators by
\begin{equation}\label{fieldop}
a_j({\bf r_\perp})=\sqrt{V d} \int \frac{d^3k}{(2\pi)^3} a_k e^{i({\bf k_\perp\cdot r_\perp}+ j k_z d)}.
\end{equation}
Analogous expressions relate $b_j({\bf r_\perp})$ and $b_k$.
Here, and in similar equations from the main text, the integral is over all $\bf k_\perp$, but $-\pi/d<k_z<\pi/d$, and $V$ is the volume of the system.  The length of the unit cell in the $z$-direction is $d$.  Substituting these relations into the Hamiltonian, yields (for $t>0$)
%
%The single particle Hamiltonian is diagonal in momentum space, and for $t>0$ can be expressed as 
%a sum of terms,
\begin{eqnarray}
H&=&V \int \frac{d^3k}{(2\pi)^3}\left(\begin{array}{cc} a_{\bf k}^{\dagger} & b_{\bf k}^{\dagger} \end{array} \right) H_k  \left( \begin{array}{c} a_{\bf k} \\ b_{\bf k} \end{array} \right)\\
H_{\bf k} &=& \left(\begin{array}{cc} \frac{\hbar^2 k_{\perp}^2}{2 m} & -2 J^{\prime} \cos(k_z d/2) \\  -2 J^{\prime} \cos(k_z d/2) & \frac{\hbar^2 k_{\perp}^2}{2 m} - \Delta \end{array}\right),
\end{eqnarray}
%where $d$ is the length of the unit cell, and 
%\begin{equation}\label{fieldop}
%a_j({\bf r_\perp})=\sqrt{V d} \int \frac{d^3k}{(2\pi)^3} a_k e^{i({\bf k_\perp\cdot r_\perp}+ j k_z d)}. 
%\end{equation}
%
In the experimentally relevant regime, $\Delta \gg J^{\prime}$, the dispersion relation for the upper and lower band respectively are given by:
\be\label{disapp}
\epsilon_{\bf k}^{H} = 
%\sqrt{(\frac{\Delta}{2})^2 + (2 J^{\prime} \cos(k_z d/2)^2}- \frac{\Delta}{2} + \frac{\hbar^2 k_{\perp}^2}{2 m},
J(1+\cos(k_z d)) + \frac{\hbar^2 k_{\perp}^2}{2 m},
\ee
\be
\epsilon_{\bf k}^{L} = -\Delta -J(1+\cos(k_z d)) + \frac{\hbar^2 k_{\perp}^2}{2 m}
\ee
where $J = 2 (J^{\prime})^2/\Delta$.  
The eigenstates for higher and lower band respectively are given by:
\begin{equation} \label{hb}
|\psi(k)\rangle^{H} = \overline{a}_{k}^{\dagger} \vert 0\rangle \approx \left(a_k^{\dagger} - \frac{2 J^\prime \cos(k_z d/2)}{\Delta} b_k^{\dagger}\right) |0\rangle.
\end{equation}
\begin{equation}\label{lb}
    |\psi(k)\rangle^{L} = \overline{b}_{k}^{\dagger} \vert 0\rangle \approx \left(b_k^{\dagger} + \frac{2 J^\prime \cos(k_z d/2)}{\Delta} a_k^{\dagger}\right) |0\rangle.
\end{equation}
Before the quench, the system is condensed in a state of the same form as EQ.~(\ref{hb}), but with $\Delta\to-\Delta$.  Since the overlap between these states are near unity, the quench projects the condensate into the higher band.

\section{Discretization}\label{discretization}
To numerically integrate Eq.~(\ref{mastereq}) and (\ref{gamma}), we discretize energy and time, using bin sizes $\delta \varepsilon$ and $\delta \tilde t$.  Integrals over $\varepsilon$ become sums, and we evaluate functions of $\varepsilon$ at the midpoint of each bin.  We used both an Euler method and a fourth order Runge-Kutta method for our time-stepping.  We chose our time-step so that the estimated temporal discretization error is at the sub-percent level.  We use $\varepsilon_{\rm max}=20$ as our largest bin, and verified that the resulting errors were on the percent level.

%The smallest energy bin requires special treatment, since the density of states vanishes at $\varepsilon=0$.  Therefore we store the total number of particles in that bin rather than the occupation factor $f$.

The temporal scaling with the number of energy bins $N_\varepsilon$ is poor, with each evaluation of the integrals in Eq.~\ref{full1} taking a time that scales as $N_\varepsilon^{3}$.  We calculate the kinetics with $\delta\varepsilon=0.1$, $0.05$, $0.025$ and $0.0125$ corresponding to $N_\varepsilon=200,400,800,1600$.  

%We denote the resulting distribution functions at midpoints of energy bins where $f_{\varepsilon} \tilde\rho(\varepsilon) d\varepsilon$ stores the number of particles within the bin of the energy spacing size around that energy. 

We use the number of atoms in our smallest energy bin as a proxy for the number of atoms condensed at $k=\pi$.  In equilibrium, this approach overestimates the number of condensed particles by a factor which scales with $\sqrt{\delta \varepsilon}$.  To correct for this factor, we run our simulation with multiple values of $\delta\varepsilon$ and extrapolate to $\delta\varepsilon\to 0$.

%Given the density of states which scale as $\sqrt{\varepsilon}$ for small $\varepsilon$ and the Bose equilibrium distribution, our lowest discrete energy bin overestimates the number of particles in the condensed zero energy state by  factor which scales with $\sqrt{d\varepsilon}$. Thus with different energy spacings, we extrapolate to the true number of particles in the zero energy state to plot the condensate fractions in our plots. 

\section{Derivation of Boltzmann Equation}\label{BTE}
Following standard arguments \cite{kadanoff}, we begin with Fermi's Golden Rule, and write the rate of change of the occupation of the mode with momentum $k$ as
%
%
%In particular, the equation of motion for the occupation of the mode with momentum $\bf k$ is
\begin{equation}\label{fgr}
\frac{\partial N_k}{\partial t} = \sum_f |\langle f |H_{\rm int} |i\rangle|^2 (N_k^f-N_k^i)
\frac{2\pi}{\hbar} \delta(E_f-E_i),
\end{equation}
where the states $|i\rangle$ and $|f\rangle$ have definite numbers of particles in each momentum state.  Here  $N_k^{i}$ and $N_k^{f}$ are the initial and final number of particles in state $k$. The energy of each state is $E_i$ and $E_f$.  

The interaction Hamiltonian involves taking particles with momentum $k_1$ and $k_2$ scatter into $k_3$ and $k_4=k_1+k_2-k_3$.  In particular, we use the interactions from Eq.~(\ref{hk}),    
\begin{equation}\label{hintap}
H_{int}=\frac{U}{2} \frac{d}{V} \sum_{k_1k_2k_3}  a_{\bf k_1+k_2-k_3}^{\dagger}  a_{\bf k_3}^{\dagger} a_{\bf k_2} a_{\bf k_1}
\end{equation}

\subsection{Explicit kinetic equations}

\subsubsection{Non-condensed Contributions}
We will first consider the terms not involving condensates -- for which   $k_1,k_2,k_3,$ and $k_4$ can be taken as distinct.  
There are four terms in Eq. (\ref{hintap}) which connect $i$ to $f$, corresponding to permuting the various indices.  The sum of these four equal contributions yields,
\begin{equation}\label{me1}
\langle f |H_{\rm int}|i\rangle = \frac{2 U d}{V} \sqrt{N_1}\sqrt{N_2}\sqrt{1+N_3}\sqrt{1+N_4},
\end{equation}
where we have used the shorthand $N_j=N_{k_j}$.
Thus the contribution to $\partial N_k/\partial t$ from these terms are
\begin{eqnarray}\label{qbe}
\frac{\partial N_k^{(1)}}{\partial t} &=& \frac{2 U^2 d^2}{\hbar}\!\!\nonumber
\int\!\!\frac{d^3 q \,d^3 k^\prime}{(2\pi)^6}  \Upsilon
\, 2\pi \delta( \epsilon_k + \epsilon_{k^\prime}-\epsilon_{k-q} - \epsilon_{k^\prime+q})\\
\Upsilon&=&
[N_{k-q} N_{k^\prime+q} (1+N_k)(1+N_{k^\prime}) \\\nonumber&&\quad- N_k N_{k^\prime} (1+ N_{k-q}) (1+ N_{k^\prime+q})]
\end{eqnarray}
%\begin{eqnarray}
%\left(\frac{\partial N_k}{\partial t}\right)_1&=&\nonumber
%\left(\frac{2 U d}{V}\right)^2 \frac{1}{2} \sum_{k_1,k_2} \Upsilon\frac{2\pi}{\hbar} \delta(\epsilon_k +\epsilon_3-\epsilon_2-\epsilon_1),\\
%\Upsilon&=&\left[(1+N_k)(1+N_3)N_2 N_1\right.\\&&\qquad\left. -N_k N_3 (1+N_1)(1+N_2)\right]\nonumber
%\end{eqnarray}
The superscript $(1)$ indicates that we have not yet included the condensate contributions.

\subsubsection{Condensate Contributions}\label{evap1}
In the presence of a condensate, we also have to separately consider terms where two atoms scatter out of the condensate, or the reverse.  
 There is no way to conserve energy and scatter two particles into or out of $k=\pi$, so we only need to worry about such terms for the condensate at $k=0$.  
 Thus we take $|f\rangle$ to differ from $|i\rangle$ by having two fewer particles with momenta $k=0$, and two more particle with momenta respectively $q$ and $-q$.  The matrix element is
\begin{equation}
\langle f |H_{\rm int}|i\rangle =  \frac{U d}{V} \sqrt{N_0-1}\sqrt{N_0}\sqrt{1+N_q}\sqrt{1+N_{-q}},
\end{equation}
Note the factor of $2$ different from Eq.~(\ref{me1}), as there are only two terms in $H_{\rm int}$ which contribute, instead of 4.  The net result is
\begin{eqnarray}\label{ev1}
\frac{\partial N_0}{\partial t} &=&\nonumber
 (Ud)^2 \frac{N_0^2}{\hbar V} \int \!\!\frac{d^3 q}{(2\pi)^3} \bar{\Upsilon}\, 2\pi \tilde\delta(2\epsilon_q-2\epsilon_0)\\\label{decayeq}
 \bar{\Upsilon}&=&  \left[ N_q N_{-q}- (1+N_q)(1+ N_{-q})\right]\\
\frac{\partial N_q^{(2)}}{\partial t} &=& \frac{U^2 d^2 N_0^2}{\hbar V^2}(1+N_q+N_{-q})\,2\pi \tilde\delta(2\epsilon_q-2\epsilon_0)\nonumber
\end{eqnarray}
where we have assumed $N_0\gg 1$. The superscript (2) indicates that we are only considering the condensate contributions. In the standard derivation of the quantum Boltzmann equation, $\tilde\delta$ is simply a Dirac delta function.  For the decay of the condensate, the finite condensate lifetime is important, so we take
\begin{equation}
2\pi\tilde\delta(2\epsilon) = \frac{2\hbar\Gamma}{(2\epsilon)^2+(\hbar\Gamma)^2}.
\end{equation}
The decay rate $\Gamma$ should be calculated self-consistently:
\begin{equation}
\Gamma = -\frac{1}{N_0} \frac{\partial N_0}{\partial t}.
\end{equation}

\subsubsection{Further Considerations -- Necessity of Self-Consistently Including the Lifetime of the Condensate Mode.}

It is crucial that the delta-function in Eq.~(\ref{ev1}) is replaced by a Loentzian -- for otherwise one gets incorrect results.  

To understand this necessity, consider solving Eq.~(\ref{ev1}) in the absence of the redistribution terms in Eq.~(\ref{qbe}). Further imagine treating $\Gamma$ as a constant, rather than self-consistently solving for it.  The standard approach of neglecting the broadening would correspond to taking the limit $\Gamma\to0$.

Under these circumstances, condensate decay only occurs into modes where $|\epsilon_q-\epsilon_0|$ is no greater than $\Gamma$.  There are roughly $V \rho_0 \Gamma$ of these, where $\rho_0=m/(\hbar^2 d)$ is the density of states per unit volume.  Thus the average occupation of a mode will be of order $N_q\sim \hbar^2 n d/(m \Gamma)$.  If this becomes larger than 1, then Bose-enhancement is important for setting the rate of decay.  In particular if $\Gamma\to 0$, the decay rate becomes infinitely fast.  This is clearly unphysical.

As already presented, the correct way to control this divergence is to find $\Gamma(t)$ self-consistently.

\subsubsection{ Limits of Validity}

If the condensate decay rate $\Gamma$ becomes large compared to the bandwidth $2J$, then quantum coherent effects need to be included:  The single particle states become strongly hybridized, and the quantum state is no longer well characterized by just specifying the occupations of different $k$ modes.  Therefore we will require $\Gamma\ll 2J$.  As discussed in the main text, this requirement means that our approach is only accurate for sufficiently large $\alpha$.

\subsection{Ergodic Approximation and adimensionalizing}\label{erg}

We make the ergodic approximation, where all states with the same energy are taken to be equally occupied: $N_k=f(\epsilon_k)$.  We convert our expressions into equations for $f(\epsilon)$ by using
\begin{equation}
\int \frac{d^{3}k}{(2\pi)^3} \frac{\partial N_k}{\partial t} 2\pi \delta(\epsilon-\epsilon_k)=
\frac{1}{V} \rho(\epsilon) \frac{\partial f(\epsilon)}{\partial t}.
\end{equation}

After making the ergodic approximation, we adimensionalize our equations.  We measure times in terms of
\begin{equation}
\tau_N= \frac{2\hbar V} {N  \rho_0 (U d)^2},
\end{equation}
denoting $\tilde t = t/\tau_N$.  For the kinetic processes in Eq.~(\ref{qbe}) we find it convenient to rescale energies by $J$, writing $\varepsilon=\epsilon/J$. We further adimensionalize momenta by rescaling, $k_z=q_z/d$, and ${\bf k_\perp}={\bf q_\perp}/ \sqrt{\hbar^2/2mJ}$.

In terms of these variables, Eq.~(\ref{qbe}) becomes 
\begin{eqnarray}\label{full1}
\frac{\partial f^{(1)}(\varepsilon_1) }{\partial \tilde t }&=& %\beta \tilde\rho(\varepsilon_1)
\frac{1}{\tilde \rho(\varepsilon_1)}
  \int\!\! \frac{d{\varepsilon_2}d{\varepsilon_3}d{\varepsilon_4}}{(2\pi)^3}
 M^{12}_{34}\Pi^{12}_{34}\bar\Delta
 \end{eqnarray}
 where energy conservation comes from
 \begin{eqnarray}
 \bar\Delta&=&
%\nonumber\\
%&&\qquad
2\pi\delta(\varepsilon_1+\varepsilon_2-\varepsilon_3-\varepsilon_4) 
\end{eqnarray}
%where $\beta = \frac{4}{\alpha (N/N^*)}$ with 
%$\alpha= \frac{N}{V} \frac{\tau_N}{\rho_0}$.  
The occupation numbers enter in the coefficient
\begin{equation}
M^{12}_{34} = f_3 f_4(1+ f_2)(1+f_1)-  f_1 f_2(1+ f_3)(1+ f_4),
\end{equation}
where $f_j=f(\varepsilon_j)$.
The dimensionless matrix element is
\begin{eqnarray}
\Pi^{12}_{34}
 &=& 
%What is right prefactor???
A
\int Dk\, \Delta_1\Delta_2\Delta_3\Delta_4  K_{1234}\label{pieq}
\end{eqnarray}
where
\begin{eqnarray}
A&=&%4 \frac{V}{N} \frac{J^2}{\rho_0^2}\left(\frac{2mJ}{\hbar^2 d}\right)^3\\
32 \frac{N^*}{0.85 N}\\
Dk&=&
\frac{d^3 q_1}
{(2\pi)^3}
\frac{d^3 q_2}{(2\pi)^3}
 \frac{d^3 q_3}{(2\pi)^3}
 \frac{d^3 q_4}{(2\pi)^3}
%\frac{d^3 k_1 d^3 k_2 d^3 k_3 d^3 k_4}{(2\pi)^{12}} 
\\
\Delta_j&=&
2\pi \delta(\varepsilon_j-\varepsilon(q_j)). \\
K_{1234}&=&
(2\pi)^3 \delta^3(q_1+q_2-q_3-q_4)\label{keq}
\end{eqnarray}
are respectively the amplitude, measure, energy conserving delta-functions, and a momentum conserving delta function.  
%In section~\ref{} we will approximate these integrals.
In appendix~\ref{MatrixElement} we approximate Eq.~\ref{pieq} as
\begin{equation}
\frac{\Pi^{12}_{34}}{A}\approx 
\frac{1}{\frac{64\pi^2}{\sqrt{\varepsilon_2}}+ \frac{2\pi (\varepsilon_3 \varepsilon_4)^{1/2}}{ K(\varepsilon_1\varepsilon_2/(\varepsilon_3\varepsilon_4))}} \end{equation}
which is exact for both high energy and low energy collisions, and is numerically efficient to calculate.

%Note, that there is only one dimensionless parameter that uniquely determines the evolution: $N/N^*$.
%
%the integrals in Eqs~(\ref{pieq}) through (\ref{keq}) are in terms of dimensionful variables, and are adimensionalized by $A$:  The characteristic energy scale is $J$ and the characteristic length scale is $(\rho_0 J)^{-1/3}$.  Up to these scalings, and coefficients of order unity, $A\sim N^*/N$.  Note that $J$ drops out of the equations expressed in this way.
%
%The factors of $N$ are sensible:  Consider the high temperature classical limit.  There $f$ is typically of order $N/V$, and  $M\sim N^2/V^2$.  Hence $f^{-1} \partial_t f \propto V/N$, as expected for two-body collisions. 

After rescaling, Eq.~(\ref{decayeq}) becomes
\begin{eqnarray}\label{nounitdecay1}
\frac{df(\varepsilon)}{d \tilde t} &=& 0.85 \frac{N}{N^*} M^2 (1+2f(\varepsilon))\frac{\bar \Gamma}{(\varepsilon-2)^2 + (\bar \Gamma/2)^2}  \\  
\bar \Gamma &=& -0.85\frac{N/N^*}{\alpha} \frac{1}{M}\frac{dM}{d \tilde t} \nonumber\\\label{nounitdecay}
&=& 0.85\frac{N/N^*}{\alpha} M \int d\varepsilon \frac{\tilde\rho(\varepsilon)(1+2f(\varepsilon))\bar \Gamma}{(\varepsilon-2)^2 + (\bar \Gamma/2)^2}   
\end{eqnarray}
where $M=N_0/N$.   
 We have assumed the condensate fraction is large, $N_0\gg1$.
%Here
 %$\tilde t = t/\tau_N$, 
 $ \bar \Gamma = \hbar\Gamma/J$ is the adimensionalized condensate evaporation rate.
 Number conservation is cast as
 \begin{equation}\label{conservation}
 M+\frac{1}{0.85 N/N^*}\int \tilde\rho(\varepsilon)f(\varepsilon)\,d\varepsilon = 1.
 \end{equation}

 \section{Dimensionless Matrix Element $\Pi_{12}^{34}$}\label{MatrixElement}
 
 Here we calculate the matrix element in Eq.~(\ref{full1}).
 
 \subsection{Low Energy Limit}
 
 We first evaluate the matrix element integral in the low energy limit where all of the energies have $\varepsilon\ll 1$.  In that case one can expand about the minimum, and it becomes a standard 3D gas calculation.  In particular, shifting the origin and using dimensionless energy and momenta,
\begin{equation}
\varepsilon(k)\approx k_\perp^2 + k_z^2.
\end{equation}
We first go to the center of mass frame in momentum for Eq.~(\ref{pieq}) to get:
\begin{eqnarray}\label{lowenergy}
\Pi^{12}_{34}&=& A 
\int \frac{d^3 K}{(2\pi)^3} \int \frac{d^3 q}{(2\pi)^3} \int \frac{d^3 q^\prime}{(2\pi)^3} \delta_1 \delta_2 \delta_3 \delta_4
\end{eqnarray}
where,
\begin{eqnarray}
q_1 = K/2 + q\\
q_2 = K/2 - q\\
q_3 = K/2 + q^\prime\\
q_4 = K/2 - q^\prime\\
\delta_1 = 2\pi\delta(\varepsilon_1 - |K/2+q|^2)\\
\delta_2 = 2\pi\delta(\varepsilon_2 - |K/2-q|^2)\\
\delta_3 = 2\pi\delta(\varepsilon_3 -|K/2+q^\prime|^2)\\
\delta_4 = 2\pi\delta(\varepsilon_4 - |K/2-q^\prime|^2)
\end{eqnarray}

Next we transform to spherical coordinates, letting $\theta$ be the angle between $K$ and $q$, and $\theta^\prime$ be the angle between $K$ and $q^\prime$. We can do the angular integrals followed by the $q$ and $q^\prime$ integrals to get:

\begin{eqnarray}\label{intermediate1}
\Pi^{12}_{34}&=& \frac{A}{16}\frac{1}{(2\pi)^2} 
\int \!\! d K \theta_1 \theta_2 \theta_3 \theta_4
\end{eqnarray}
where,
\begin{eqnarray}\nonumber
\theta_1 &=& \theta\left( \left[\frac{\varepsilon_1+\varepsilon_2}{2} - \frac{K^2}{4} \right] K^2-\left[\frac{\varepsilon_1-\varepsilon_2}{2}  \right]^2\right)\\\nonumber
\theta_2 &=&\theta\left( \left[\frac{\varepsilon_3+\varepsilon_4}{2} - \frac{K^2}{4} \right] K^2-\left[\frac{\varepsilon_3-\varepsilon_4}{2}  \right]^2\right)\\\nonumber
\theta_3 &=& \theta\left(\varepsilon_1+\varepsilon_2-K^2/2\right)\\
\theta_4 &=& \theta\left(\varepsilon_3+\varepsilon_4-K^2/2\right)
\end{eqnarray}
where throughout $\theta(x)$ is the Heaviside step function (equal to 1 when $x>0$ and otherwise zero).
The integrand in Eq.~(\ref{intermediate1}) is always zero or 1.  The latter occurs when 
\begin{eqnarray}
 \label{in1}
|\sqrt{\varepsilon_1}-\sqrt{\varepsilon_2}|<K<\sqrt{\varepsilon_1}+\sqrt{\varepsilon_2}\\
\label{in2}
|\sqrt{\varepsilon_3}-\sqrt{\varepsilon_4}|<K<\sqrt{\varepsilon_3}+\sqrt{\varepsilon_4} \\
\label{in3}
K<\sqrt{2}\sqrt{\varepsilon_1+\varepsilon_2}\\
\label{in4}
K<\sqrt{2}\sqrt{\varepsilon_3+\varepsilon_4}.
\end{eqnarray}
It is convenient to write 
\begin{eqnarray}
\varepsilon_1&=&\bar\varepsilon+\delta\\
\varepsilon_2&=&\bar\varepsilon-\delta\\
\varepsilon_3&=&\bar\varepsilon+\delta^\prime\\
\varepsilon_4&=&\bar\varepsilon-\delta^\prime.
\end{eqnarray}
Let us further assume that $\delta>\delta^\prime>0$.  That means that $\epsilon_2<\epsilon_4<\epsilon_3<\epsilon_1$.
Consequently 
\begin{eqnarray}
|\sqrt{\varepsilon_1}-\sqrt{\varepsilon_2}|^2&=& 2\bar\varepsilon - \sqrt{\bar\varepsilon^2-\delta^2}\\
&>& 2\bar\varepsilon - \sqrt{\bar\varepsilon^2-(\delta^\prime)^2}\\
&=& |\sqrt{\varepsilon_3}-\sqrt{\varepsilon_4}|^2
\end{eqnarray}
and
\begin{eqnarray}
|\sqrt{\varepsilon_1}+\sqrt{\varepsilon_2}|^2&=& 2\bar\varepsilon + \sqrt{\bar\varepsilon^2-\delta^2}\\
&<& 2\bar\varepsilon + \sqrt{\bar\varepsilon^2-(\delta^\prime)^2}\\
&=& |\sqrt{\varepsilon_3}+\sqrt{\varepsilon_4}|^2,
\end{eqnarray}
Hence the integral is just
\begin{eqnarray}
\Pi^{12}_{34}&=& \frac{A}{16}\frac{1}{(2\pi)^2} \sqrt{\varepsilon_2}.
\end{eqnarray}
Of course, this result was predicated on $\epsilon_2$ being the smallest energy.  More generally we have
\begin{eqnarray}
\Pi^{12}_{34}&=& \frac{A}{16}\frac{1}{(2\pi)^2} 
 {\rm Min}(\sqrt{\varepsilon_1},\sqrt{\varepsilon_2},\sqrt{\varepsilon_3},\sqrt{\varepsilon_4}).
\end{eqnarray}
This is a well-known classic result in kinetic theory \cite{landau}.  

%This has the right dimensions [1/(length$^9$ energy$^4$].

\subsection{High Energy}
Next we consider the case where all of the $\varepsilon$'s are large compared to $1$.  We can then approximate
\begin{equation}
\varepsilon(k) \approx k_\perp^2,
\end{equation}
and neglect the $k_z$ dependence. All momenta here are dimensionless.  We do the $k_z$ integrals and scale and recenter the momenta as in Eq.~(\ref{lowenergy}) to arrive at
\begin{eqnarray}
\Pi^{12}_{34}&=& A
 \int \frac{d^2K}{(2\pi)^2} \int \frac{d^2q}{(2\pi)^2} \int \frac{d^2q^\prime}{(2\pi)^2} \delta_1 \delta_2\delta_3\delta_4
 \end{eqnarray}
 %where,
 %\begin{eqnarray}
 %\delta_1 = 2\pi\delta(\varepsilon_1-|K/2+q|^2)\\
 %\delta_2 = 2\pi \delta(\varepsilon_2-|K/2-q|^2)\\ 
% \delta_3 = 2\pi\delta(\varepsilon_3-|K/2+q^\prime|^2)\\
% \delta_4 = 2\pi\delta(\varepsilon_4-|K/2-q^\prime|^2)
%\end{eqnarray}
We transform to polar coordinates, letting $\theta$ be the angle between $K$ and $q$, and $\theta^\prime$ be the angle between $K$ and $q^\prime$. Doing the angular integral first, followed by the integral over $q$ and $q^\prime$, we get:

\begin{eqnarray}\label{intermediate2}
\Pi^{12}_{34}&=&
 \frac{A}{16}\int \frac{d(K^2)}{4\pi}\frac{1}{\sqrt{f(K)}}\frac{1}{\sqrt{g(K)}}
 \end{eqnarray}
 where,
\begin{eqnarray}
f(K) = \left(\frac{\varepsilon_1+\varepsilon_2}{2} -\frac{K^2}{4}\right)K^2-\left(\frac{\varepsilon_1-\varepsilon_2}{2}\right)^2\\
g(K) = \left(\frac{\varepsilon_3+\varepsilon_4}{2} -\frac{K^2}{4}\right)K^2-\left(\frac{\varepsilon_3-\varepsilon_4}{2}\right)^2
\end{eqnarray}
Here the integral is taken over the domain where the arguments of the square roots are positive.  We know from our previous arguments that if we take $\varepsilon_2<\varepsilon_4<\varepsilon_3<\varepsilon_1$ then $K_{\rm min}=\sqrt{\varepsilon_1}-\sqrt{\varepsilon_2}$ and $K_{\rm max}=\sqrt{\varepsilon_1}+\sqrt{\varepsilon_2}$, or $K_{\rm min}^2=\varepsilon_1+\varepsilon_2-2\sqrt{\varepsilon_1 \varepsilon_2}$ and $K_{\rm max}^2=\sqrt{\varepsilon_1}+\sqrt{\varepsilon_2}+ 2 \sqrt{\varepsilon_1\varepsilon_2}$.  

Equation~\ref{intermediate2} is an elliptic integral.  To show that, we factor the expressions in the square roots, to get
\begin{eqnarray}
\Pi^{12}_{34} = \frac{A}{16\pi}
 \int dK^2 \frac{1}{\sqrt{p_1 p_2 p_3 p_4}}
 \end{eqnarray}
 where,
 \begin{eqnarray}
 p_1 = \left(K^2-2\bar\varepsilon-2\sqrt{\varepsilon_1\varepsilon_2}\right)\\
p_2 = \left(K^2-2\bar\varepsilon+2\sqrt{\varepsilon_1\varepsilon_2}\right)\\
p_3 = \left(K^2-2\bar\varepsilon-2\sqrt{\varepsilon_3\varepsilon_4}\right)\\
p_4 = \left(K^2-2\bar\varepsilon-2\sqrt{\varepsilon_3\varepsilon_4}\right)
\end{eqnarray}
where $\bar\varepsilon=(\varepsilon_1+\varepsilon_2)/2=(\varepsilon_3+\varepsilon_4)/2$.  We then shift and rescale $K^2$, writing
\begin{equation}
s=\frac{K^2-2\bar\varepsilon}{2\sqrt{\varepsilon_1\varepsilon_2}}
\end{equation}
to find
\begin{eqnarray}
\Pi^{12}_{34}&=&
\frac{A}{4\pi}
\frac{1}{\sqrt{\varepsilon_1\varepsilon_2}} \int_{-1}^1 \frac{ds}{\sqrt{ \left(s^2-1\right)\left(s^2-\frac{\varepsilon_2\varepsilon_3}{\varepsilon_1\varepsilon_2}\right)}}.
\end{eqnarray}
This is the Jacobi notation for the complete Elliptic Integral of the first kind,
\begin{equation}
K(1/t)=\frac{\sqrt{t}}{2}\int_{-1}^1 \frac{ds}{ \sqrt{(s^2-1)(s^2-t)}},
\end{equation}
which gives
\begin{eqnarray}
\Pi^{12}_{34}&=&
\frac{A}{2\pi}
\frac{1}{\sqrt{\varepsilon_3\varepsilon_4}} K\left(\frac{\varepsilon_1\varepsilon_2}{\varepsilon_3\varepsilon_4}\right).
\end{eqnarray}
By construction $\varepsilon_1\varepsilon_2<\varepsilon_3\varepsilon_4$. 
More generally,
\begin{eqnarray}
\Pi^{12}_{34}&=&
\frac{A}{2\pi}
\frac{1}{\sqrt{E_2}} K\left(\frac{E_1}{E_2}\right),
\end{eqnarray}
where $E_1=\min(\varepsilon_1 \varepsilon_2,\varepsilon_3 \varepsilon_4)$, and  $E_2=\max(\varepsilon_1 \varepsilon_2,\varepsilon_3 \varepsilon_4)$.
% If not, as in Eq.~(\ref{}), we reverse them.
%
%Again the dimensional analysis works out.

\subsection{Interpolation}
To connect these two limits 
we use a simple interpolation 
\begin{eqnarray}
\Pi_{12}^{34} = \frac{A}{\frac{64\pi^2}{\sqrt{\varepsilon_2}}+ \frac{2\pi (\varepsilon_3 \varepsilon_4)^{1/2}}{ K(\varepsilon_1\varepsilon_2/(\varepsilon_3\varepsilon_4))}}.
\end{eqnarray}
This is exact in both limits.

\end{document}